\documentstyle[12pt,aps]{revtex}
\def\sigcom{\sigma_{\pi{\scriptscriptstyle N}}}

\def\MN{M_{{\scriptscriptstyle N}}}
\def\half{{\textstyle{1\over 2}}}
\def\third{{\textstyle{1\over 3}}}
\def\dslash{\partial\llap/}
\def\psibar{\overline\psi}

\def\mbar{\overline m}

\begin{document}

\title{
\begin{flushright}
\vspace{-1cm}
{\normalsize MC/TH 96/08}
\vspace{1cm}
\end{flushright}
Chiral symmetry in nuclei\footnote{Talk presented at the Nuclear Physics Study
Weekend on Extreme States of Nuclei, Abingdon, October 1995. A more complete 
review of the ideas presented here can be found in [1] and an introduction for
nonspecialists in [2].}\bigskip}
\author{Michael C. Birse}
\address{Theoretical Physics Group, Department of Physics and Astronomy,\\
University of Manchester, Manchester, M13 9PL, U.K.\\}
\maketitle
\vskip 20pt
\begin{abstract}
The role of the quark condensate in low-energy QCD and its behaviour in
nuclear matter are discussed. Partial restoration of chiral symmetry, as
indicated by a reduction of the quark condensate in matter, could significantly
alter the properties of nucleons and mesons. Various signals of these effects
are discussed: enhancement of the axial charge of a nucleon, lepton pair
production and $\overline K$-nucleus attraction.
\end{abstract}

\section{Introduction}

Chiral symmetries are features of theories with massless fermions, where the
fields describing right- and left-handed particles decouple. They are
preserved by interactions with vector fields, such as photons, gluons, $W$
and $Z$ bosons. In contrast, interactions with a Lorentz scalar character,
such as mass terms, couple right- and left-handed fields and so break these 
symmetries. 

The importance of chiral symmetries has been recognised since the observation
of parity violation in the weak interaction: when a nucleus $\beta$-decays
it emits a (dominantly) left-handed electron and a right-handed antineutrino.
The interaction couples only to left-handed particles and right-handed
antiparticles and is invariant under a left-handed weak isospin symmetry. 
The interactions of quarks and leptons thus respect symmetries of massless
particles, yet the particles themselves clearly have masses. The answer to
this apparent conundrum was provided by Nambu: the chiral symmetries are
hidden (or ``spontaneously broken"). By this we mean that, although the
dynamical equations describing the particles respect these symmetries, the
ground state or vacuum does not.\footnote{For more on the symmetries of
the ``Standard Model" of the strong, electromagnetic and weak interactions
and their dynamical consequences, see\cite{dgh}.}

The interactions of quarks and leptons with this asymmetric vacuum give rise
to the masses of these particles. An important consequence of this is that
their masses should not be regarded as universal constants, but can depend
on the environment of the particles.

In the case of quarks we should be careful to distinguish two contributions to
their masses. The current mass of a quark is generated by the Higgs field of
the electroweak interaction, and so is similar in origin to the lepton masses.
The current masses range from around 7 MeV for the up and down quarks to 
$\sim\! 180$ GeV for the top quark. In contrast, the dynamical mass of a quark
is produced by the strong interaction. It arises because the vacuum  of Quantum
Chromodynamics (QCD) contains a condensate of quark-antiquark pairs, rather
like the condensate of Cooper pairs of electrons in a superconductor. For the
up and down quarks, these masses are around 300--400 MeV and they are the
relevant ones for simple quark models of mesons and baryons.

The typical energy scale associated with the Higgs field is $\sim\! 100$ GeV, 
corresponding to the temperature at which a phase transition is expected to
a state of massless quarks, leptons, $W$ and $Z$ bosons. Hence for all 
practical purposes we can regard these current masses of quarks as constants.

In contrast, the typical energy scale of QCD is $\sim\! 100$ MeV, the
temperature at which the transition to a quark-gluon plasma is expected. Of
the various phase transitions that occurred as the early universe cooled down
after the big bang, this is the only one that we may be able to study
experimentally, using ultrarelativistic heavy-ion collisions\cite{jones}.

A similar transition to quark matter is also expected at low temperature but
high density. This may be relevant to the extreme states of matter at the
centres of neutron (or quark?) stars. Although these densities are hard to
create in the laboratory, precursors of this transition may already be present
at normal nuclear densities, as discussed here.

\section{Chiral symmetry}

\subsection{SU(2)$\times$SU(2)}

As already mentioned the current masses of the up and down quarks are very much
smaller that typical hadronic masses. Hence to a very good approximation the
QCD Lagrangian is invariant under both ordinary isospin rotations,
$$\psi\rightarrow(1-\half i\hbox{\boldmath$\beta$}\cdot\hbox{\boldmath$\tau$})
\psi, \eqno(2.1)$$
and axial isospin rotations, 
$$\psi\rightarrow(1-\half i\hbox{\boldmath$\alpha$}\cdot\hbox{\boldmath$\tau$}
\gamma_5)\psi,\eqno(2.2)$$
where \hbox{\boldmath${\alpha}$} and \hbox{\boldmath${\beta}$} denote
infinitesimal parameters. By taking combinations of these rotations involving
$1\pm\gamma_5$ we can independently rotate the isospin of right- and
left-handed massless quarks. This chiral symmetry is thus known as
SU(2)$_R\times$SU(2)$_L$.

Interactions with vector fields (such as gluons and photons) respect this
symmetry since the current $\psibar\gamma_\mu\psi$ is invariant under axial
rotations. In contrast, the scalar and pseudoscalar densities of quarks are
not invariant, transforming under (2.2) as
$$\psibar\psi\rightarrow\psibar\psi-\hbox{\boldmath$\alpha$}\cdot\psibar i
\hbox{\boldmath$\tau$}\gamma_5\psi,$$
$$\psibar i\hbox{\boldmath$\tau$}\gamma_5\psi\rightarrow\psibar i
\hbox{\boldmath$\tau$}\gamma_5\psi+\hbox{\boldmath$\alpha$}\psibar\psi.
\eqno(2.3)$$
Hence fermion mass terms or couplings to scalar fields break the symmetry. 

The Noether currents corresponding to the transformations (2.1,2) are the
(vector) isospin current 
$${\bf V}^\mu=\psibar\gamma^\mu\half\hbox{\boldmath$\tau$}\psi, \eqno(2.4)$$
and the axial current
$${\bf A}^\mu=\psibar\gamma^\mu\gamma_5\half\hbox{\boldmath$\tau$}\psi. 
\eqno(2.5)$$

Although SU(2)$_R\times$SU(2)$_L$ is a good approximate symmetry of the QCD
Lagrangian, it is hidden by the asymmetric nature of the vacuum. As a result
pions are very close to being the massless ``Goldstone bosons" of this 
symmetry, which explains why they are so much lighter than other hadrons. An
order parameter that describes this aspect of the vacuum is the nonzero
expectation value of the scalar density of quarks, a quantity often called
``the quark condensate." 

Another consequence of the noninvariant vacuum is the matrix element that
allows charged pions to decay via the axial-current part of weak interaction:
$$\langle 0|A^\mu_i(x)|\pi_j(q)\rangle= i f_\pi q^\mu e^{-i q\cdot x} 
\delta_{ij}, \eqno(2.6)$$
where the pion decay constant is $f_\pi=92.4\pm 0.3$ MeV\cite{pdg}. 
The divergence of this equation,
$$\langle 0|\partial_\mu A^\mu_i(x)|\pi_j(q)\rangle=f_\pi m_\pi^2 
e^{-i q\cdot x} \delta_{ij}, \eqno(2.7)$$
allows us define ``interpolating" pion fields,
$$\hbox{\boldmath$\phi(x)$}={1\over f_\pi m_\pi^2}\partial_\mu{\bf A}^\mu(x)
\eqno(2.8)$$
These operators connect the vacuum and one-pion states with the same
normalisation that canonical pion fields would have. The advantage of this
choice is that, by going to the soft-pion limit $q\rightarrow 0$, we can
relate amplitudes for pion scattering to the axial transformation properties of
the states involved.

These ideas can be extended to include the three lightest flavours of quarks,
although the resulting symmetry is more strongly broken by the larger current
mass of the strange quark and hence kaons are less close to being Goldstone
bosons.

\subsection{Linear sigma model}

A convenient example of a field theory that embodies the basic features of
chiral symmetry is the linear sigma model\cite{gml}. This based on a chiral
multiplet of meson fields $\sigma$, {\boldmath$\phi$} that transform under
SU(2)$_R\times$SU(2)$_L$ like the quark bilinears $\psibar\psi$ and
$\psibar i\hbox{\boldmath$\tau$}\gamma_5\psi$ in (2.3). Its Lagrangian has the
form
$${\cal L}=\psibar\bigl[i\dslash+g(\sigma+i\hbox{\boldmath$\phi$}\cdot
\hbox{\boldmath$\tau$}\gamma_5)\bigr]\psi+\half(\partial_\mu\sigma)^2
+\half(\partial_\mu\hbox{\boldmath$\phi$})^2-U(\sigma,\hbox{\boldmath$\phi$}),
\eqno(2.9)$$
where the fermion fields $\psi$ could describe either nucleons or quarks, 
depending how we want to use the model.

The potential energy has a symmetric ``Mexican hat" form, 
$$U_0(\sigma,\hbox{\boldmath$\phi$})={\lambda^2\over 4}\big(\sigma^2+
\hbox{\boldmath$\phi$}^2-f_\pi^2\big)^2, \eqno(2.10)$$
familiar in many other areas of physics, such as ferromagnets and the Higgs
sector of electroweak theory. The physical vacuum $\sigma=-f_\pi$,
$\hbox{\boldmath $\phi$}=0$ lies in the circular minimum running round the
brim of the hat. It is not chirally invariant and so the symmetry is hidden.
With this potential, pions are massless Goldstone bosons while the $\sigma$
mesons are massive. The crown of the hat, where the fields vanish, is the state
where manifest chiral symmetry is restored. It becomes the minimum-energy
configuration in the presence of matter at high enough temperature or density.

The model can be made more realistic by adding an explicit symmetry breaking
term:
$$U(\sigma,\hbox{\boldmath$\phi$})=U_0(\sigma,\hbox{\boldmath$\phi$})
+f_\pi m_\pi^2\sigma. \eqno(2.11)$$
This tips the Mexican hat, breaking the symmetry and giving the pions their
observed masses.

\section{Quark condensate}

\subsection{Gell-Mann--Oakes-Renner}

As an example of PCAC, consider the pion propagator in vacuum, defined using
the pion fields of (2.8): 
$$\third\sum_i\int\!d^4\!x\,e^{i q\cdot x}\langle 0|{\rm T}\bigl(
\partial_\mu A^\mu_i(x),\partial_\nu A^\nu_i(0)\bigr)|0\rangle=i{f_\pi^2 
m_\pi^4\over q^2-m_\pi^2}f(q^2),\eqno(2.12)$$
where we know the strength with which our fields couple to pions on their mass
shell and so $f(m_\pi^2)=1$. The dynamical assumption of PCAC is that
$f(q^2)$ is a smoothly varying function (and $m_\pi$ is small on typical
hadronic scales). Hence to a good approximation we can replace $f(0)$ by its
on-shell value value when we extrapolate off-shell to $q^2=0$. In this
``soft-pion" limit, we can integrate (2.12) by parts and rewrite it as
$$\third\sum_i\langle 0|\bigl[Q_5^i,[Q_5^i,{\cal H}(0)]\bigr]|0\rangle
\simeq-f_\pi^2 m_\pi^2. \eqno(2.13)$$
where the $Q_5^i$ are the axial charge operators and ${\cal H}(x)$ is the
Hamiltonian density. This double commutator picks out the part of the
Hamiltonian that breaks the symmetry and so (2.14) gives a connection between
the pion mass and the strength of the symmetry breaking, known as a
Gell-Mann--Oakes--Renner (GOR) relation \cite{gor}. Its form may also be
recognised as that of an energy-weighted sum rule saturated by a single state,
the pion.

The symmetry-breaking part of the QCD Hamiltonian is $\mbar\psibar\psi$ and
so the GOR relation can be written as
$$\mbar\langle 0|\psibar\psi|0\rangle\simeq-f_\pi^2 m_\pi^2. \eqno(2.14)$$
Chiral symmetry thus provides a link between a QCD matrix element on the 
l.h.s.~and pion observables on the r.h.s. 

The current masses of the light quarks have been estimated from hadron mass
splittings and QCD sum rules\cite{qmass}, although these methods are not very
precise. Values for $\mbar$ lie in the range 5--10 MeV (for a renormalisation
scale of 1 GeV). Since $\mbar$ is not known to within a factor of two, the
quark condensate is similarly uncertain. Typical values for $\langle
0|\psibar\psi|0\rangle$ are around $-3$ fm$^{-3}$. Note the negative sign,
which implies that the positive scalar densities of quarks in hadrons will
tend to reduce the magnitude of the condensate in matter.

\subsection{Sigma commutator}

To discuss what happens to the quark condensate in nuclear matter, we also need
the corresponding matrix element for a single nucleon. This is known as the
pion-nucleon sigma commutator and can be written in a form analogous to (2.13),
$$\sigcom=\third\sum_i\langle N|\left[Q_5^i,[Q_5^i,H]\right]
|N\rangle,  \eqno(2.15)$$
where $|N\rangle$ denotes a zero-momentum nucleon state. The commutator
is thus the contribution of chiral symmetry breaking to the nucleon mass. It 
can also be expressed in terms of the integrated scalar density of quarks in 
the nucleon as
$$\sigcom=\mbar\langle N|\int\!d^3\!{\bf r}\,\psibar\psi|N\rangle.
\eqno(2.16)$$

A value for $\sigcom$ can be deduced from $\pi N$ scattering amplitudes by 
using dispersion relations and extrapolating to the soft-pion limit. The most
recent determination by Gasser {\it et al.}\ gives $\sigcom=45\pm7$ MeV
\cite{gls}, with a significant uncertainty because of inconsistencies between
the data sets used in the extrapolation. Taking a typical value of $\mbar=7$
MeV for the current quark mass, we find that this corresponds to an integrated
scalar density of quarks in the nucleon of about 6, much larger than the 3 one
would expect from a simple quark model.

This difference from simple quark models arises from the cloud of virtual
pions that surrounds the nucleon. These can be important for many nucleon
properties, but are especially so for $\sigcom$. Here the cloud can contribute
20--25 MeV\cite{jtc,bm1}. Added to the 12--25 MeV from the valence quarks,
this can explain the value deduced from experiment. Moreover the importance of
the pion cloud means that the form factor associated with $\sigcom$ is long
ranged, with a radius of about 1.3 fm. Again this is consistent with the 
results of Gasser {\it et al.}\cite{gls}.

\section{Quark condensate in matter}

Having presented values for the scalar densities of quarks in the vacuum and a
single nucleon, I turn now to nuclear matter. At low densities we can treat the
nucleons as independent and simply add their contributions to get the spatially
averaged scalar density of quarks: 
$$\mbar\langle \psibar\psi\rangle_\rho=\mbar \langle \psibar\psi\rangle_0
+\sigcom\,\rho, \eqno(3.1)$$
where $\rho$ is the density of nucleons. This leading (linear) density
dependence is independent of any model for the interactions between the
nucleons\cite{dl,cfg,lkw}. The poorly known current mass can be cancelled by
taking the ratio to the vacuum condensate and using the GOR relation (2.14) to
get
$${\langle \psibar\psi \rangle_\rho\over \langle \psibar\psi\rangle_0}
=1-{\sigcom\over f_\pi^2m_\pi^2}\rho. \eqno(3.2)$$
If we assume that this linear density dependence holds up to nuclear matter 
density, $\rho_0\simeq 0.17$ fm$^{-3}$, then we expect a $\sim\! 30$\% reduction 
in the quark condensate in nuclear matter. In terms of the picture provided by
the linear sigma model, the vacuum has been pushed towards the crown of the
Mexican hat. Chiral symmetry is thus partially restored, which could have a
variety of interesting consequences, such as decreases in the constituent
masses of quarks (and hence of hadron masses).

This result for the condensate in matter raises three questions:
\begin{itemize}
\item What are the corrections to it from higher-order dependence on the 
density?
\item How do correlations between nucleons affect the degree of symmetry
restoration?
\item What are the consequences of partial symmetry restoration for nucleon and
meson properties? 
\end{itemize}

\subsection{Corrections}

Simple arguments suggest that terms of higher order in $\rho$ should be small.
As suggested by Cohen {\it et al.}\cite{cfg}, the Feynman-Hellmann theorem (a
consequence of the variational principle) can be used to estimate the scalar 
density from models for nuclear matter. Applying this to the dependence on 
$\mbar$ of the energy density of nuclear matter ${\cal E}$ gives
$$\mbar\langle\psibar\psi\rangle=\mbar{d{\cal E}\over d\mbar}. \eqno(3.3)$$
This energy can be written in the form
$${\cal E}={\cal E}_0+\MN\,\rho+B(\rho), \eqno(3.4)$$
where ${\cal E}_0$ is the energy density of the normal vacuum, the term linear
in $\rho$ arises from the rest masses of the nucleons, and $B(\rho)$ includes
all terms of higher order in $\rho$, arising from the nucleons' kinetic and
potential energies. Hence the quark condensate can be expressed as
$$\mbar\langle\psibar\psi\rangle_\rho=-f_\pi^2m_\pi^2+\sigcom\,\rho
+\Delta\sigma(\rho), \eqno(3.5)$$
where three terms correspond to the terms in (3.4), and in particular
$\Delta\sigma(\rho)$ arises from the binding energy. Since the binding energy
per nucleon is less than 2\% of $\MN$ in nuclear matter, it would be rather
remarkable if the higher-order corrections in $\Delta\sigma(\rho)$ were not
much smaller than the linear term.

This expectation is borne out by estimates of the density dependence of the
quark condensate in various models of nuclear matter. Contributions from pion 
exchange show a cancellation between Pauli blocking of the pion cloud and pion
exchange with tensor correlations\cite{ce}. Similarly heavier-meson exchanges
indicate that there are strong cancellations between scalar mesons, which
contribute attractively in the $NN$ interaction, and vector mesons, which are
repulsive\cite{cfg,bm2}. Estimates using realistic $NN$ interactions in
relativistic BHF calculations confirm this\cite{lk,bw}, showing very small
deviations from linear density dependence up to the density of normal nuclear
matter. Beyond this, higher-order effects set in rapidly and the extrapolations
become very model-dependent. Clearly more reliable models are needed for the
study of nuclear matter at very high densities.

\subsection{Correlations}

In examining the consequences of the change in the quark condensate, the range
of the effect is crucial. If it were short-ranged then the spatial average
of $\langle \overline\psi\psi\rangle$ would not be a useful quantity to
consider. The strong repulsive correlations between nucleons would leave them
as isolated islands of restored symmetry surrounded by a sea of normal 
vacuum\cite{te}.

However such a picture is not realised since, as described above, the pion
cloud provides a large part of the scalar density of a single nucleon. The
scalar density can thus extend well beyond the quark core of that nucleon. The
corresponding contributions to the symmetry restoration experienced by a
second nucleon can be expressed in terms of two-pion exchange between the
nucleons. The partial symmetry restoration is thus long-ranged, and should
survive in the presence of hard-core correlations\cite{bir}.

Calculations in the linear sigma model\cite{bir} suggest a connection between
chiral symmetry restoration and the attractive central interaction between
nucleons. Both arise from exchange of two pions with zero net spin and isospin
and have a similar range. However the various diagrams appear in different
combinations and so, while closely related, symmetry restoration and the
central attraction are not identical.

\subsection{Consequences}

Partial restoration of chiral symmetry reduces the dynamical quark mass and 
so it is also expected to reduce the masses of nucleons and mesons in nuclear
matter. In the case of nucleons this is very similar to the role of the scalar
fields of Dirac phenomenology\cite{dirac}. A rule of thumb suggested by the
simple scaling arguments of Brown and Rho\cite{brown} is that the decrease in
hadron masses is similar to that in the condensate:
$${M_N^*\over M_N}\sim {m_v^*\over m_v} \sim {\langle 
\overline\psi\psi\rangle_\rho\over \langle \overline\psi\psi\rangle_0}, 
\eqno(3.6)$$
where the stars indicate values for nucleon and vector meson masses in matter.

The fact that these changes act at the quark level means that the internal
structure of the nucleon can also be modified. The consequences for nucleon
properties in medium have been estimated in a variety of models
(see\cite{birrev} for references). Most of these predict qualitatively similar
effects: increases in the proton charge radius and nucleon magnetic moments
and a decrease in the coupling to the axial current. There have also been
suggestions that these modifications of nucleon structure could explain the
EMC effect: differences between the momentum distribution of quarks in a
nucleus compared with that for a free nucleon\cite{emc}. Combined with changes
in meson masses these modifications of nucleons could alter the effective 
interaction between nucleons in matter, for example reducing the net tensor
force. 

However with all of these possible signals, it is hard to disentangle changes
in nucleon structure from more conventional many-body effects. There are thus
very few observables that provide clean evidence for changes in nucleon
properties.

\section{Signals of symmetry restoration}

\subsection{Axial charge}

One exception, which does seem to provide a clear indication of strong scalar
fields in nuclei, is the effective axial charge. This is the matrix element of
the time-component of the axial current. To leading order in a nonrelativistic
expansion, the one-nucleon part of this operator is
$$\hbox{\bf A}_0=g_A^c{\hbox{\boldmath $\sigma\cdot${\bf p}}\over \MN}
\hbox{\boldmath $\tau$}, \eqno(4.1)$$
and so is proportional to the nucleon velocity. For a free nucleon the axial
charge coupling is equal to the more familiar axial current coupling,
$g_A^c=g_A=1.257\pm 0.003$. In nuclei an effective axial charge is defined by
$$\langle f|\hbox{\bf A}_0|i\rangle={g_A^{c*}\over \MN}\langle f|
(\hbox{\boldmath $\sigma\cdot${\bf p}})\hbox{\boldmath $\tau$}|i\rangle_{sm},
\eqno(4.2)$$
where $\langle\cdots\rangle_{sm}$ denotes the corresponding shell-model matrix
element. All effects of correlations, meson exchange currents and nucleon
modifications are contained in the effective charge $g_A^{c*}$.

Interest in this quantity was reawakened by Warburton's studies of
first-forbidden $\beta$-decays of nuclei in the lead region\cite{warb},
showing enhancements of $\sim\! 80$--100\% in the effective axial charge.
One-pion exchange has long been known to make a significant contribution to
this quantity\cite{kdr}, but this enhancement is limited to at most 
$\sim\! 50$\% by a soft-pion theorem. Relativistic treatments show that the
extra enhancement needed to explain the data can be provided by exchanges of
heavier mesons\cite{krtt}.

Of particular importance is the exchange of a scalar, isoscalar meson. In
phenomenological $NN$ interactions this $\sigma$ meson represents the exchange 
of pairs of pions and is responsible for the intermediate-range attraction in
relativistic models of nuclei\cite{dirac}. At the mean-field level this
interaction reduces the nucleon mass. Hence the velocity of a nucleon of given
momentum increases, as does its contribution to the axial charge:
$$g_A^{c*}=g_A{\MN\over\MN^*}. \eqno(4.3)$$
Note that, although in Dirac treatments of nucleons this effect arises from 
``Z-graphs", the form of this contribution to the enhancement is general, being
determined by relativistic invariance alone. A similar contribution appears 
when nucleons are treated as composite particles\cite{birac}.

Support for this as evidence of scalar fields in nuclei is provided by the
the reaction $pp\rightarrow pp\pi^0$ close to threshold, which probes similar
physics and whose cross section shows a large enhancement over that 
expected from the one-nucleon process\cite{meyer,lr}.

The enhancement of the axial charge thus provides strong evidence for large
scalar fields in nuclei and a corresponding reduction in the nucleon mass.

\subsection{Vector mesons}

In nuclear matter, vector meson masses are also expected to
decrease\cite{hlak}. The electromagnetic decays of these mesons into $e^+e^-$
pairs provide a possible way to ``see" inside dense matter in way that cannot
be done with strongly interacting probes. Two types of experiment are planned:
at CEBAF E-94-002 will use photoproduction of vector mesons on nuclei, while
at GSI the HADES detector will study lepton pairs from hot dense matter
produced in relativistic heavy-ion collisions at around 1 GeV per nucleon.

An intriguing hint of what these experiments may find is provided by recent
results from the CERES collaboration on $e^+e^-$ pairs from S on Au collisions
at 200 GeV per nucleon\cite{ceres}. Compared with $pp$ collisions, an
enhancement is seen in the production of pairs with invariant masses around
400 MeV (between the $\pi^0\rightarrow e^+e^-\gamma$ Dalitz decay and $\rho$
peaks). The spectrum of pairs has been rather well described\cite{lkb} by a
relativistic-transport model in which the $\rho$-meson mass is reduced to 380
MeV in the initial fireball ($\rho\simeq 3.5\rho_0$ and $T\simeq 185$ MeV).
However one should bear in mind that the uncertainties in the data are large.
Also an alternative explanation has been proposed\cite{crw}, based on medium
effects on the $\pi\pi$ channel to which the $\rho$ couples strongly.

\subsection{Pions and kaons}

Pions and kaons, as approximate Goldstone bosons, are rather different from
heavier mesons. Their nonzero masses arise from the explicit breaking of
chiral symmetry, as can be seen from the GOR relation (2.14). However, as that
relation also shows, their masses depend on the quark condensate as well.

In order to estimate the changes in the pion mass, one needs to know how both 
this condensate and the decay constant change in matter. At least at low 
densities, where the impulse approximation holds, one can relate these to pion
scattering from a single nucleon. The amplitude for this contains an
energy-independent term provided by the sigma commutator (or quark condensate)
and an energy-dependent one, which corresponds to the change in the decay
constant in matter. At threshold these cancel almost exactly to leave a very
small isospin-symmetric scattering length. One therefore expects the mass of 
a low-momentum pion to be almost unchanged in matter at low
densities\cite{dee}. If the quark condensate were to vanish at some high
density, one might find interesting effects\cite{cb}. Either chiral symmetry
could be restored or, if $f_\pi$ did not vanish at the same point, the pion
could become very light, possibly signalling the onset of $s$-wave pion
condensation\cite{bkr}.

The behaviour of kaons in matter is less well understood but a similar
mechanism could lead to $s$-wave kaon condensation in dense matter. The
original estimates of this\cite{kn} assumed an unnaturally large $KN$ sigma
commutator but more realistic values still give a significant (scalar)
attraction for both kaons and antikaons. Combined with the vector interaction,
which is attractive for antikaons, this could lead to $K^-$ condensation at a
few times the density of nuclear matter\cite{bkrt}. That would have important
consequences for supernovae and the formation of neutron stars \cite{bb}.

At low densities however the impulse approximation leads to a weakly repulsive
$K^-$-nucleus potential, as a consequence of the $\Lambda(1405)$ resonance
just below the $K^-p$ threshold. Models that predict $K^-$ condensation
suggest that at higher densities this resonance is moved above the $K^-p$
threshold so that the potential then becomes attractive\cite{lbr}. Some support
for this picture is provided by recent fits to data on kaonic atoms which find
a density-dependent optical potential that is weakly repulsive in the nuclear
surface and strongly attractive in the interior\cite{fgb}.

A similar attraction could also affect kaons in the hot dense matter produced
in relativistic heavy-ion collisions\cite{ko}, and this could provide a
mechanism for the strangeness enhancement seen in the NA36
experiment\cite{na36,jones} that does not involve a quark-gluon plasma.

\section{Conclusions}

The model-independent result (3.2) for the quark condensate suggests that
there is a significant degree of chiral symmetry restoration inside nuclei.
Corrections to this result have been estimated and shown to be small up to
normal nuclear densities. However extrapolations to higher densities show
large and strongly model-dependent corrections. The symmetry restoration can
have a longish range, similar to that of the attractive central potential
between nucleons. Its effects should thus not be cut out by the hard-core
correlations between nucleons. 

Consequences of a partial restoration of chiral symmetry are often hard to
disentangle for more conventional many-body effects. One exception is the
enhancement of the axial charge which provides evidence for strong scalar
fields in nuclei. Further evidence may come from studies of vector mesons
in matter using their decays into lepton pairs.

\section*{Acknowledgments}

This work is supported by the EPSRC and PPARC. I am grateful to J.~McGovern
for collaboration on the topics described here and for many helpful 
discussions.


\begin{references} 
\bibitem{birrev}M. C. Birse, J. Phys.\ G: Nucl.\ Part.\ Phys.\ {\bf 20} (1994)
1537.
\bibitem{pworld}M. C. Birse and J. A. McGovern, Physics World (October 1995) 
p.~35.
\bibitem{dgh}J. F. Donoghue, E. Golowich and B. R. Holstein, {\it Dynamics of
the standard model} (Cambridge University Press, Cambridge, 1992).
\bibitem{jones}P. G. Jones, contribution to this meeting.
\bibitem{pdg}Particle Data Group, Phys.\ Rev.\ {\bf D50} (1994) 1443.
\bibitem{gml}M. Gell-Mann and M. L\'evy, Nuovo Cim.\ {\bf16} 1960 705
\bibitem{gor}M. Gell-Mann, R. Oakes and B. Renner, Phys.\ Rev.\ {\bf 175}
(1968) 2195.
\bibitem{qmass}J. Gasser and H. Leutwyler, Phys.\ Report {\bf 87} (1982) 77.
\bibitem{gls}J. Gasser, H. Leutwyler and M. E. Sainio, Phys.\ Lett.\ {\bf B253}
(1991) 252, 260.
\bibitem{jtc}I. Jameson, A. W. Thomas and G. Chanfray, J. Phys.~G 
{\bf 18} (1992) L159.
\bibitem{bm1}M. C. Birse and J. A. McGovern, Phys.\ Lett.\ {\bf B292} (1992)
242. 
\bibitem{dl}E. G. Drukarev and E. M. Levin, Nucl.\ Phys.\ {\bf A511} (1990)
679; {\bf A516} (1990) 715(E). 
\bibitem{cfg}T. D. Cohen, R. J. Furnstahl and D. K. Griegel, Phys.\ Rev.\ {\bf
C45} (1992) 1881. 
\bibitem{lkw}M. Lutz, S. Klimt and W. Weise, Nucl.\ Phys.\ {\bf A542} (1992) 
521.
\bibitem{bm2}M. C. Birse and J. A. McGovern, Phys.\ Lett.\ {\bf B309}
(1993) 231. 
\bibitem{ce}G. Chanfray and M. Ericson, Nucl.\ Phys.\ {\bf A56} (1993) 427.
\bibitem{lk}G. Q. Li and C. M. Ko, Phys.\ Lett.\ {\bf 338} (1994) 118.
\bibitem{bw}R. Brockmann and W. Weise, Phys.\ Lett.\ {\bf B367} (1996) 40.
\bibitem{te}T. E. O. Ericson, Phys.\ Lett.\ {\bf B321} (1994) 312.
\bibitem{bir}M. C. Birse, Phys.\ Rev.\ {\bf C49} (1994) 2212.
\bibitem{brown}C. Adami and G. E. Brown, Phys. Reports {\bf 224} (1993) 1;\\
G. E. Brown and M. Rho, hep-ph/9504250, Phys. Reports (to appear).
\bibitem{dirac}B. D. Serot and J. D. Walecka, Adv. Nucl. Phys. {\bf 16} 
(1985) 327;\\
S. J. Wallace, Annu. Rev. Nucl. Part. Sci. {\bf 37} (1987) 267.
\bibitem{emc}J. J. Aubert {\it et al.}, Phys.\ Lett.\ {\bf B123} (1983) 275;\\
L. Frankfurt and M. Strickman, Phys.\ Reports {\bf 160} (1988) 235;\\
R. P. Bickerstaff and A. W. Thomas, J. Phys.\ G: Nucl.\ Part.\ Phys. {\bf 15} 
(1989) 1523;\\
M. Arneodo, Phys.\ Reports {\bf 240} (1994) 301, and references
therein.
\bibitem{warb}E. K. Warburton, Phys.\ Rev.\ Lett.\ {\bf 66} (1991) 1823;
Phys.\ Rev.\ {\bf C44} 233;\\
E. K. Warburton and I. S. Towner, Phys.\ Lett.\ {\bf B294} (1992) 1.
\bibitem{kdr}K. Kubodera, J. Delorme and M. Rho, Phys.\ Rev.\ Lett.\ {\bf 40} 
(1978) 755.
\bibitem{krtt}M. Kirchbach, D. O. Riska and K. Tsushima, Nucl.\ Phys.\ {\bf 
A542} (1992) 616;\\
I. S. Towner, Nucl.\ Phys.\ {\bf A542} (1992) 631;\\
E. K. Warburton, I. S. Towner and B. A. Brown, Phys.\ Rev.\ {\bf C49} (1994) 
824.
\bibitem{birac}M. C. Birse, Phys.\ Lett.\ {\bf B316} (1993) 472; 
Phys.\ Rev.\ {\bf C51} (1995) R1083.
\bibitem{meyer}H. O. Meyer {\it et al.}, Nucl.\ Phys.\ {\bf A539} (1992) 633.
\bibitem{lr}T.-S. H. Lee and D. O. Riska, Phys.\ Rev.\ Lett.\ {\bf 70} (1993) 
2237.
\bibitem{hlak}T. Hatsuda and S. H. Lee, Phys.\ Rev.\ {\bf C46} (1992) R34;\\
M. Asakawa and C. M. Ko, Phys.\ Rev.\ {\bf C48} (1993) R526.
\bibitem{ceres}G. Agakichiev {\it et al.}, Phys.\ Rev.\ Lett.\ {\bf 75} 
(1995) 1272.
\bibitem{lkb}G. Q. Li, C. M. Ko and G. E. Brown, Phys.\ Rev.\ Lett.\ {\bf 75} 
(1995) 1272.
\bibitem{crw}G. Chanfray, R. Rapp and J. Wambach, Phys.\ Rev.\ Lett.\ {\bf 76} 
(1996) 368;\\
C. Song, V. Koch, S. H. Lee and C. M. Ko, Phys.\ Lett.\ {\bf B366} (1996) 379.
\bibitem{dee}J. Delorme, M. Ericson and T. E. O. Ericson, Phys.\ Lett.\ {\bf 
B291} (1992) 379.
\bibitem{cb}T. D. Cohen and W. Broniowski, Phys.\ Lett.\ {\bf B342} (1995) 25;
{\bf B348} 12.
\bibitem{bkr}G. E. Brown, V. Koch and M. Rho, Nucl.\ Phys.\ {\bf A535} (1991) 
701.
\bibitem{kn}D. B. Kaplan and A. E. Nelson, Phys.\ Lett.\ {\bf 175} (1986) 57; 
{\bf B179} 409(E); {\bf 192} (1987) 193.
\bibitem{bkrt}G. E. Brown, K. Kubodera, M. Rho and V. Thorsson, 
Phys.\ Lett.\ {\bf 291} (1992) 355.
\bibitem{bb}G. E. Brown and H. A. Bethe, Astrophys.\ J. {\bf 423} (1994) 659.
\bibitem{lbr}C.-H. Lee, G. E. Brown and M. Rho, Phys.\ Lett.\ {\bf B335}
(1994) 266;\\
V. Koch, Phys.\ Lett.\ {\bf B337} (1995) 7;\\
C.-H. Lee, G. E. Brown, D.-P. Min and M. Rho, Nucl.\ Phys.\ {\bf A585} (1995) 
401;\\
T. Waas, N. Kaiser and W. Weise, Phys.\ Lett.\ {\bf B365} (1996) 12.
\bibitem{fgb}E. Friedman, A. Gal and C. J. Batty, Phys.\ Lett.\ {\bf B308} 
(1993) 6; Nucl.\ Phys.\ {\bf A579} (1994) 518.
\bibitem{ko}G. Q. Li, C. M. Ko and X. S. Fang, Phys.\ Lett.\ {\bf B329} (1994) 
149;\\
G. Q. Li, C. M. Ko and B. A. Li, Phys.\ Rev.\ Lett.\ {\bf 74} (1995) 235;\\
G. Q. Li and C. M. Ko, Nucl.\ Phys.\ {\bf A594} (1995) 460;\\
V. Koch, Phys.\ Lett.\ {\bf B351} (1995) 29;\\
G. Q. Li and C. M. Ko, Phys.\ Lett.\ {\bf B351} (1995) 37.
\bibitem{na36}E. Andersen {\it et al.}, Phys.\ Lett.\ {\bf B327} (1994) 433;
Nucl.\ Phys.\ {\bf A590} (1995) 291c.


\end{references}
\end{document}